\documentclass{article}

\usepackage{epsfig,amssymb,amsmath,graphicx,lineno}
\def\gsim{\mathrel\, {\vcenter {\baselineskip 0pt \kern 0pt\hbox{$>$} \kern 0pt \hbox{$\sim$} }}\,}

\author{Diego Harari, Silvia Mollerach and Esteban Roulet\\
{\it CONICET, Centro At\'omico Bariloche,}\\
{\it Av. Bustillo 9500 (8400) Argentina.}}

\title{Angular distribution of cosmic rays from an individual source in a turbulent magnetic field}

\begin{document}
\maketitle

\begin{abstract}
We obtain the angular distribution of the  cosmic rays  reaching an observer from an individual source and after propagation through a turbulent magnetic field, for different ratios between the source distance and the diffusion length. We study both the high-energy quasi-rectilinear regime as well as the transition towards the diffusive regime at lower energies where the deflections become large.  We consider the impact of energy losses, showing that they tend to enhance the anisotropy of the source at a given energy. We also discuss lensing effects, in particular those that could result from the regular galactic magnetic field component, and show that the effect of the turbulent extragalactic magnetic fields can smooth out the divergent magnification peaks that would result for point-like sources in the limit of no turbulent deflections.

\end{abstract}

\section{Introduction}
The search of the sources of the cosmic rays (CRs) is made difficult by the fact that, being charged particles, their trajectories are bent by intervening magnetic fields. Since the deflections decrease for increasing energies there is the hope that at the highest energies observed one could start to identify localized excess fluxes associated to the individual CR sources. However,  the CR arrival directions  have shown to be remarkably  isotropic at the highest energies \cite{auhs}. On the other hand,  some hints of anisotropies have been recently reported by the Pierre Auger Observatory \cite{audip} favoring the presence of a dipolar modulation with an amplitude of few percent in the CR flux at energies in excess of 8~EeV, and also some indications  have been found at the highest energies ($E\geq 60$~EeV) of localized excesses at intermediate angular scales ($10^\circ$--$20^\circ$) by both the Telescope Array \cite{tahs} and the Auger observatories \cite{auhs}. These results and the lack of small scale anisotropies suggest that the magnetic fields (intergalactic and galactic) have a non-negligible effect on the trajectories of the CRs coming from the extragalactic sources.

In ref.~\cite{I}, hereafter paper I, we computed the expected dipolar modulation of the distribution of arrival directions  of CRs after accounting for the diffusion through the intergalactic magnetic fields, and in ref.~\cite{ha15} we extended these results to the case of heavy nuclei. 
In this work we obtain the detailed angular distribution of the CRs diffusing from a source and discuss the transition between the regime of quasi-rectilinear propagation present at high rigidities in which the source  images have an approximately Gaussian spread, with an energy dependent angular width, and the more isotropic distribution appearing at lower rigidities in which the dipolar behavior is recovered and for which we also obtain the quadrupolar term of the distribution. The results depend essentially on the ratio between the distance to the source and the diffusion length. We also obtain the predictions for a superposition of several sources, both in the case of CR protons and for heavier nuclei, computing the impact of energy losses due to redshift effects and due to interactions with the radiation backgrounds.

Note that one expects that a random magnetic field could exist in the Local Supercluster with typical root mean square strength $B\simeq O(1)$~nG and coherence length $l_c\simeq 1$~Mpc. The typical deflection of a CR with energy $E$ and charge $eZ$, after propagation over a distance $L$ through a random magnetic field,  is 
\begin{equation}
\delta\simeq \frac{BZe}{E}\sqrt{\frac{Ll_c}{2}}\simeq 1^\circ \frac{B}{\rm nG}\frac{\rm 40\,EeV}{E/Z}\frac{\sqrt{Ll_c}}{\rm Mpc}.
\label{delta}
\end{equation}
 Deflections in the extragalactic magnetic field after propagating $L\gg 1$~Mpc should thus be dominant with respect to the deflection induced by the random component of the galactic magnetic field, for which $B\simeq \text{few}\ \mu\text{G}$, $l_c\simeq 10$--30~pc and $L\simeq 1$~kpc, and hence we will neglect in this work the effects of the turbulent galactic magnetic field.  On the other hand, coherent deflections produced by the regular magnetic field can be significant and in particular they can lead to sizable lensing effects due to which the CR fluxes can be strongly magnified for certain values of the energy.
In Section~\ref{lensing} we will also consider how the diffusion through the extragalactic turbulent fields can affect the predictions of the lensing effects produced by the regular galactic magnetic field.

\section{Cosmic ray angular distribution: from the ballistic to the diffusive regime}

When cosmic rays diffuse in a turbulent magnetic field there are two asymptotic regimes, depending on the relation between the distance from the source $r_s$ and the diffusion length $l_D$, which is the distance after which the deflection of the CR trajectory becomes $\sim 1$~rad. If $r_s\ll l_D$ the trajectories are quasirectilinear and CRs arrive concentrated around the direction towards the source with a Gaussian spread that increases with decreasing energy.
On the other extreme, when $r_s\gg l_D$ the CRs  perform several turns before reaching the observer and their distribution will be mostly isotropic, with the main feature being a dipole component in the direction towards the source.

In order to obtain a detailed description of the angular distribution of the CRs reaching  an observer from a given source, we perform a numerical integration of the stochastic differential equation that describes the scattering of ultrahigh-energy cosmic rays (UHECRs) in a turbulent homogeneous and isotropic magnetic field  \cite{ac99}, which is
\begin{equation}
{\rm d}n_i =-\frac{1}{l_D}n_i c{\rm d}t + \frac{1}{\sqrt{l_D}} P_{ij}{\rm  d}W_j,
\label{dni}
\end{equation}
where $P_{ij} \equiv (\delta_{ij} - n_i n_j)$ is the projection tensor onto the plane orthogonal to the direction of the CR velocity given by $\hat n\equiv (n_1,n_2,n_3)$,
repeated indices are summed and (${\rm d}W_1,\,{\rm d} W_2,\, {\rm d}W_3$) are three Wiener processes such that $\langle{\rm d} W_i \rangle =0$ and  $\langle {\rm d}W_i{\rm d}\,W_j \rangle =c\,{\rm d}t\,\delta_{ij}$. The diffusion length is  $l_D\equiv 3D/c$, where the diffusion coefficient $D$ depends on the particle rigidity and the turbulence spectrum. A fit to this dependence was obtained in  ref.\ \cite{I}, resulting in
\begin{equation}
D= \frac{c}{3}l_c\left[4\left(\frac{E}{E_c}\right)^2+a_I\left(\frac{E}{E_c}\right)+a_L\left(\frac{E}{E_c}\right)^{2-m}\right],
\end{equation}
where $E_c$ is the energy for which the effective Larmor radius $r_L\equiv E/ZeB$ equals the coherence length, i.e. $E_c=eZBl_c\simeq 0.9 Z(B/{\rm nG})(l_c/{\rm Mpc})$~EeV.
This critical energy separates the low energy regime of resonant diffusion from the high-energy quasi-rectilinear propagation regime. The index $m$ and the coefficients $a_I$ and $a_L$ depend on the distribution of magnetic energy density on different length scales. For a Kolmogorov turbulence spectrum one has  $m=5/3$, $a_L=0.23$ and $a_I=0.9$.

Considering in this section the simplified case in which energy losses are neglected,  we follow the evolution of the unit vector $\hat n$ through a numerical integration of the  stochastic differential equation
(\ref{dni}) in order to obtain the angular distribution of CRs at different distances from the source. We follow the trajectories for a sufficiently large time, and each time they cross spherical caps of different radius $r$ around the initial point we compute and record the angle $\theta$ between the instantaneous cosmic ray velocity and the direction from the source towards the CR position\footnote{Note that $\delta$ in Eq.~(\ref{delta}) refers instead to the change in the CR velocity direction between the one at the source and that at the instantaneous CR location. In the limit of small deflections one has $\langle\delta^2\rangle=3\langle\theta^2\rangle$.}. We do this for a large set of propagated particles and for different values of the quantity $R\equiv r_s/l_D$, which is the distance from the source normalized to the diffusion length.

\begin{figure}[ht]
\centerline{\epsfig{width=3.in,angle=-90,file=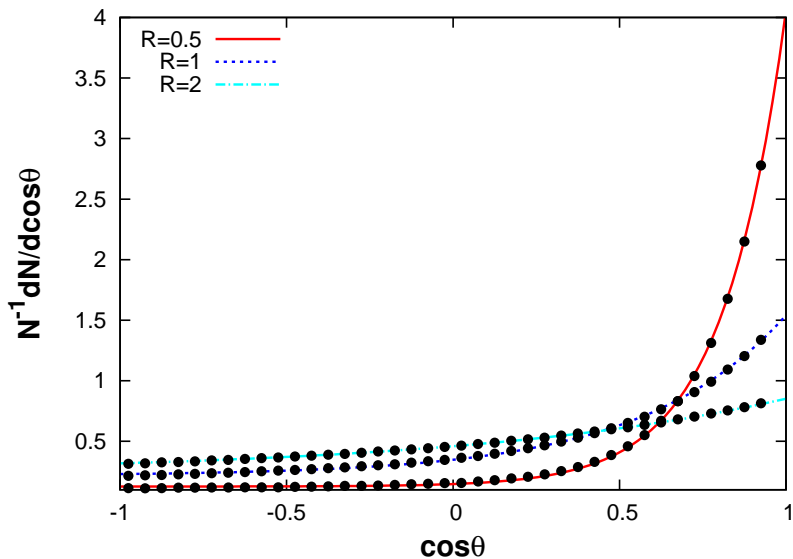}}
\caption{Distribution of CR arrival directions as a function of $\cos\theta$ for three values of the normalized source distance $R\equiv r_s/l_D=cr_s/3D=0.5,1,$  and 2. The lines indicate the fits using the expression in  Eq.~(\ref{fisher}).} 
\label{fig:f}
\end{figure}
For illustration we show in Figure $\ref{fig:f}$  the  distribution of the CR arrival directions in the cases $R=0.5,1,$ and $2$, along with the analytic fits obtained using the expression  
\begin{equation}
\frac{1}{N}\frac{\rm{d}N}{\rm{d}\cos\theta}=\frac{i}{2}+(1-i)\frac{\kappa \exp(\kappa\cos\theta)}{2\sinh\kappa}.
\label{fisher}
\end{equation}
The parameter $i$ describes a constant isotropic term, which arises mostly from particles that diffused long times and made several turns before reaching the observer. The second term follows a Fisher distribution  \cite{fi87}, which provides the natural generalization to the sphere of the Gaussian distribution and in which the `concentration' parameter $\kappa$ determines the steepness of the profile around the source direction, which becomes more pronounced  for increasing $\kappa$. It is mostly due, for $\kappa>1$, to particles that reached the observer without making any significant detour, although for $\kappa<1$ it contributes significantly to all angles $\theta$, and hence also accounts for particles which suffered large deflections. In the regime $\kappa\ll 1$ the distribution is almost isotropic, with dipole and quadrupole terms of amplitude $d=(1-i)\,\kappa$ and $q=(1-i)\,\kappa^2/2$ respectively (where d$N/\text{d}\cos\theta\simeq (N/2) (1+d\cos\theta+q(\cos^2\theta-1/3)+...)$). Note that in this limit the monopole term turns out to be actually independent from the parameter $i$, because the contribution to the isotropic part coming from the term involving the Fisher distribution, which is $(1-i)/2$, adds to the purely isotropic term $i/2$ in Eq.~(\ref{fisher}). Hence, the parameters $i$ and $\kappa$ get determined by the sizes of the dipolar and quadrupolar amplitudes.

The lines in Figure~\ref{fig:f}, which are based on Eq.~(\ref{fisher}), provide good fits to the distributions for arrival directions both in the regime of diffusion and in that of quasirectilinear propagation. For the cases illustrated the best fit parameters are $\kappa=5.4, 2.3, 1 $, and $i=0.25, 0.44, 0.47$ for $R=0.5$, 1 and 2 respectively.

The distribution of CR arrival directions can be characterized either in terms of $i$ and $\kappa$ or alternatively in terms of the first two moments $\langle\cos\theta\rangle$ and $\langle\cos^2\theta\rangle$, which are the quantities that can be obtained more directly from the simulations, through the relations
\begin{equation}
\langle\cos\theta\rangle=(1-i)\left(\coth\kappa-\frac{1}{\kappa}\right)
\label{cosq}
\end{equation}
\begin{equation}
\langle\cos^2\theta\rangle=\frac{i}{3}+(1-i)\left(1+\frac{2}{\kappa^2}-\frac{2}{\kappa \tanh\kappa}\right)=
1-\frac{2\langle\cos\theta\rangle}{\kappa}-\frac{2 i}{3}.
\label{cos2}
\end{equation}

Conversely, from these equations one finds that the parameter  $\kappa$ is related to the first two moments through
\begin{equation}
\frac{2}{3(\coth\kappa-1/\kappa)}-\frac{2}{\kappa}=\frac{\langle\cos^2\theta\rangle-1/3}{\langle\cos\theta\rangle}\equiv \alpha.
\end{equation}
The solution to this equation is accurately approximated by
\begin{equation}
\kappa\simeq \frac{5\alpha-\frac{27}{4}\alpha^2+\frac{27}{8}\alpha^3}{\frac{2}{3}-\alpha}.
\label{rk}
\end{equation}
Knowledge of $\kappa$ allows one to derive $i$ through 
\begin{equation}
i=1-\frac{\langle\cos\theta\rangle}{\coth\kappa-1/\kappa}.
\label{i}
\end{equation}

\begin{figure}[ht]
\centerline{\epsfig{width=2.in,angle=-90,file=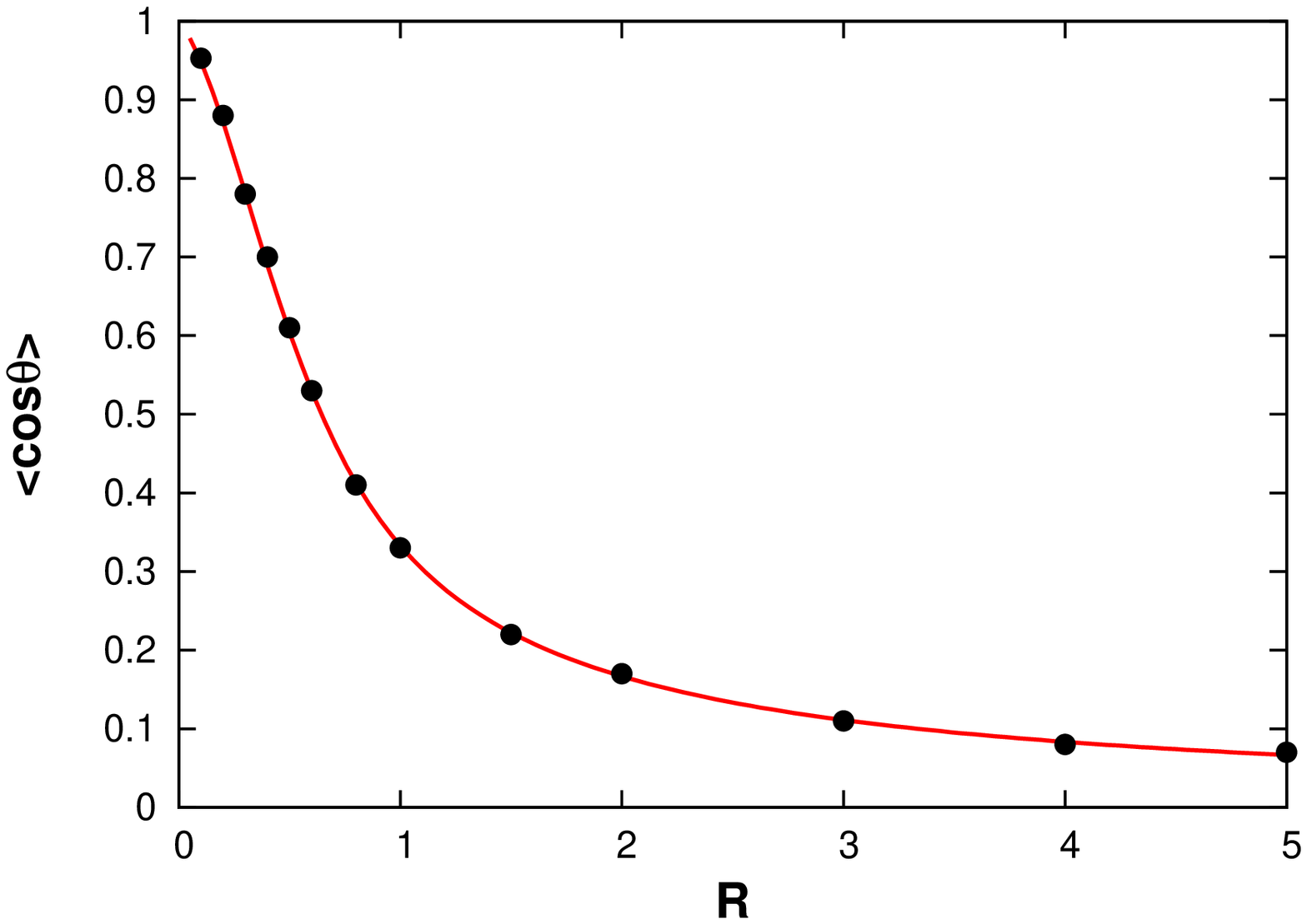}\epsfig{width=2.in,angle=-90,file=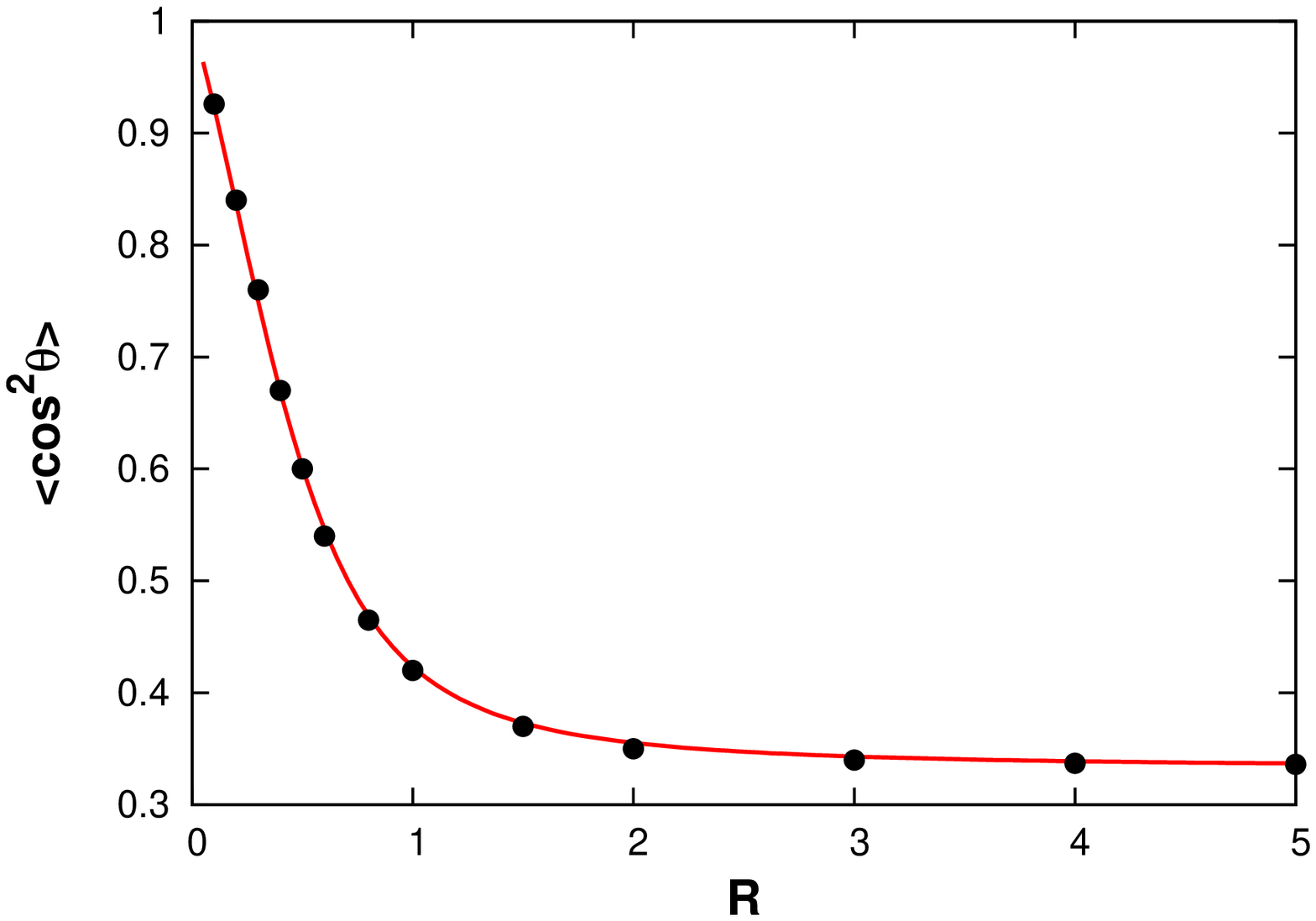}}
\caption{Average $\cos\theta$ (left) and $\cos^2\theta$ (right) as a function of the normalized source distance $R\equiv r_s/l_D=cr_s/3D$.} 
\label{fig:cos}
\end{figure}

We have shown in Paper I that the dependence of $\langle\cos\theta\rangle$ with $R\equiv r_s/l_D=cr_s/3D$ is accurately fitted by
\begin{equation}
\langle\cos\theta\rangle=\frac{1}{3R}\left[1-\exp\left(-3R-\frac{7}{2}R^2\right)\right].
\label{cos}
\end{equation}
Figure \ref{fig:cos} shows this fit (solid line) along with the results from numerical solutions of the stochastic equation. In paper I we also showed that this fit agrees with the results obtained after solving for the trajectories of charged particles in a turbulent magnetic field by a numerical integration of the Lorentz equation.

\begin{figure}[ht]
\centerline{\epsfig{width=2.in,angle=-90,file=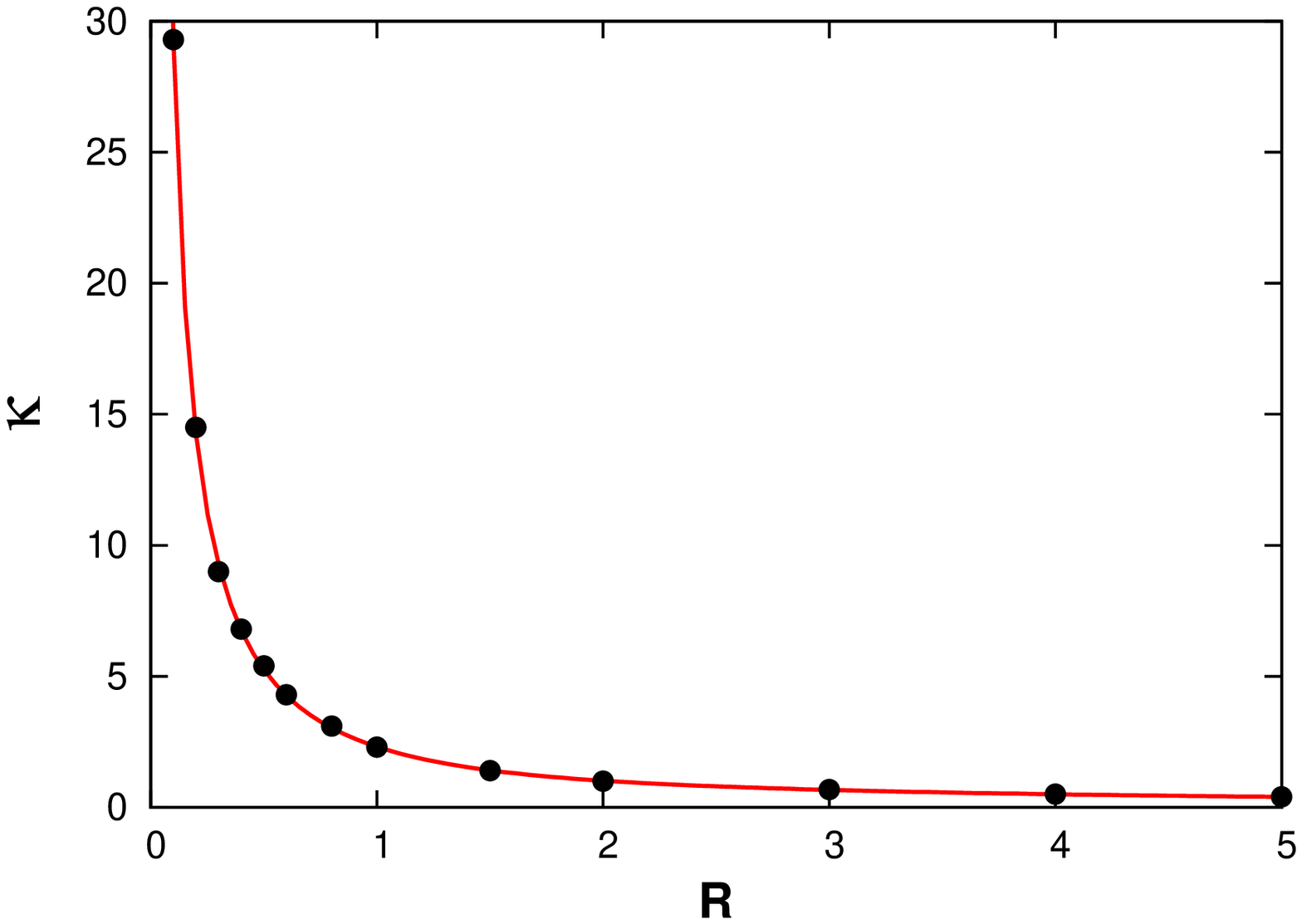}\epsfig{width=2.in,angle=-90,file=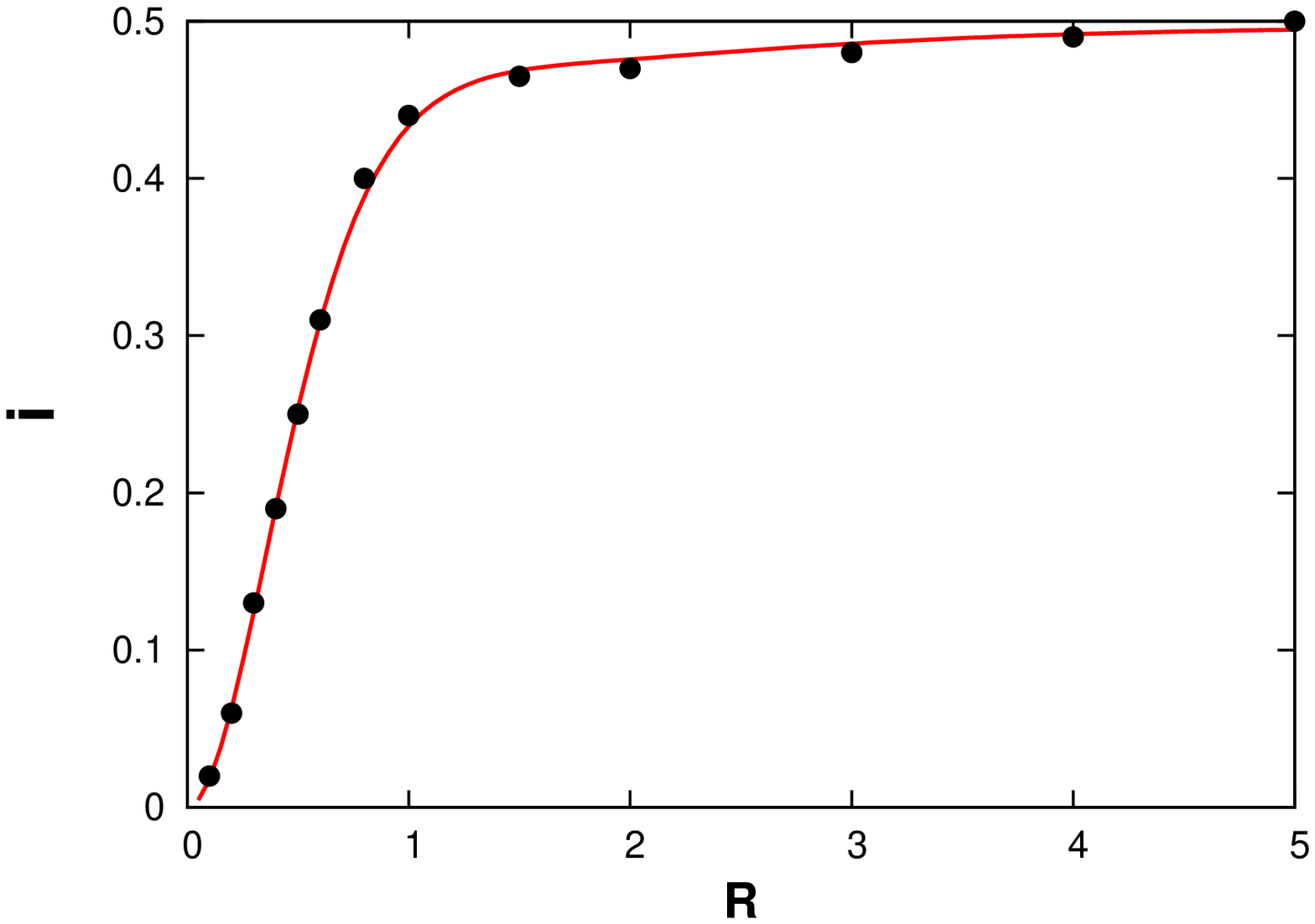}}
\vskip 1.0 truecm
\caption{Parameters $\kappa$ (left) and $i$ (right) as a function of the normalized source distance.} 
\label{fig:ki}
\end{figure}

Figure $\ref{fig:ki}$ shows as black dots the values of $\kappa$ and $i$ that best fit the angular distributions obtained through numerical integration of the stochastic equation (\ref{dni}) for several values of $R$ between 0.1 and 5.  
The solid line in the left panel is an analytic fit to the numerical results, given by
\begin{equation}
\kappa=\frac{1}{R}\left[2+\exp\left(-\frac{2}{3}R-\frac{1}{2}R^2\right)\right].
\label{k}
\end{equation}
The solid line for $i$  in the right panel of Figure $\ref{fig:ki}$ corresponds to Eq.~(\ref{i}), using $\langle\cos\theta\rangle$ and $\kappa$ as given by Eqs.~(\ref{cos}) and (\ref{k}) respectively. Hence, in the case being discussed here in which there are no energy losses, the distribution can be obtained just in terms of the relative distance to the source $R$.

In the limit  $R\ll 1$, which corresponds to the quasi-rectilinear regime,  the angular distribution is (as expected) approximately Gaussian with 
$\langle\theta^2\rangle\approx 2/\kappa \simeq 2R/3\simeq(r_s/6l_c)(E_c/E)^2$ for high energies. In the opposite limit in which $R \gg 1$  the angular distribution is well fitted with $\kappa\simeq 2/R$ and $i\simeq 0.5$, and corresponds to an almost isotropic distribution with dipolar and quadrupolar terms of amplitude $1/R$ and $1/R^2$ respectively.  

As an additional consistency check we show in the right panel of Figure \ref{fig:cos} the analytic expectation for $\langle\cos^2\theta\rangle$ (solid line) as derived using the analytic fits to $\kappa$ and $\langle\cos\theta\rangle$  along with the numerical results (dots).

\section{Angular distribution including energy losses}

The previous approach can be extended to take into account the effects of the expansion of the universe and the energy losses due to interactions with the Cosmic Microwave Background (CMB)  radiation or the extragalactic background light, as it was discussed in Paper I for the case of protons and in ref.~\cite{ha15} for the case of nuclei. 

We assume first that the cosmic rays are protons and backtrack them from the observation time at $z=0$ by numerically integrating the differential stochastic equation~(\ref{dni}). We consider a Friedmann-Robertson-Walker universe, for which 
\begin{equation}
\frac{{\rm d}z}{{\rm d}t} = H_0 (1+z) \sqrt{\Omega_m (1+z)^3+\Omega_\Lambda}
\end{equation}
with $H_0=70\ {\rm km\ s^{-1}Mpc^{-1}}$, $\Omega_m=0.3$ and $\Omega_\Lambda=0.7$. Due to energy losses from the expansion and from interactions with the CMB, we have to take into account that the particles arriving at $z=0$ with energy $E(0)$ had a larger energy $E_g(z)$ at each previous step. This effect is accounted for following the method in Paper I.

In order to obtain the distribution of arrival directions of particles originating from a source at a comoving distance $r_s$, for any given arrival energy $E(0)$ we backtrack the trajectories of a large number of particles.  We record the cosine of the angle $\theta$ between the initial velocity and the direction towards the CR position each time that the particles pass at a comoving distance $r_s$ from the original point, and we also record the energy of the particles at that time. When reconstructing the distribution of the arriving particles around the source direction we have to include a weight for each particle equal to $[E_g(z)/E(0)]^{-\gamma} {\rm d}E_g/{\rm d}E$. The first factor takes into account that for a source emitting cosmic rays with a spectrum $\propto E^{-\gamma}$ there will be less particles with the higher energy $E_g(z)$ which is required to be generated at redshift $z$ in order to reach the observer with energy $E(0)$. The second factor takes into account the change of the energy bin width from the emission to the observation. In the simulations we have considered a turbulent magnetic field with a present day ($z=0$) amplitude $B(0)=1$~nG, a coherence length $l_c(0)=1$~Mpc and a Kolmogorov spectrum. The critical energy for these parameter values is $E_c(0)= 0.9$~EeV. The coherence length scales as $l_c(z)=l_c(0)/(1+z)$ and we also assume that the magnetic field scales as $B(z)=B(0)(1+z)^{2-\mu}$, with $\mu=1$ as in \cite{be07}, in which case the critical energy is actually independent of the redshift.

\begin{figure}[ht]
\centerline{\epsfig{width=2.in,angle=-90,file=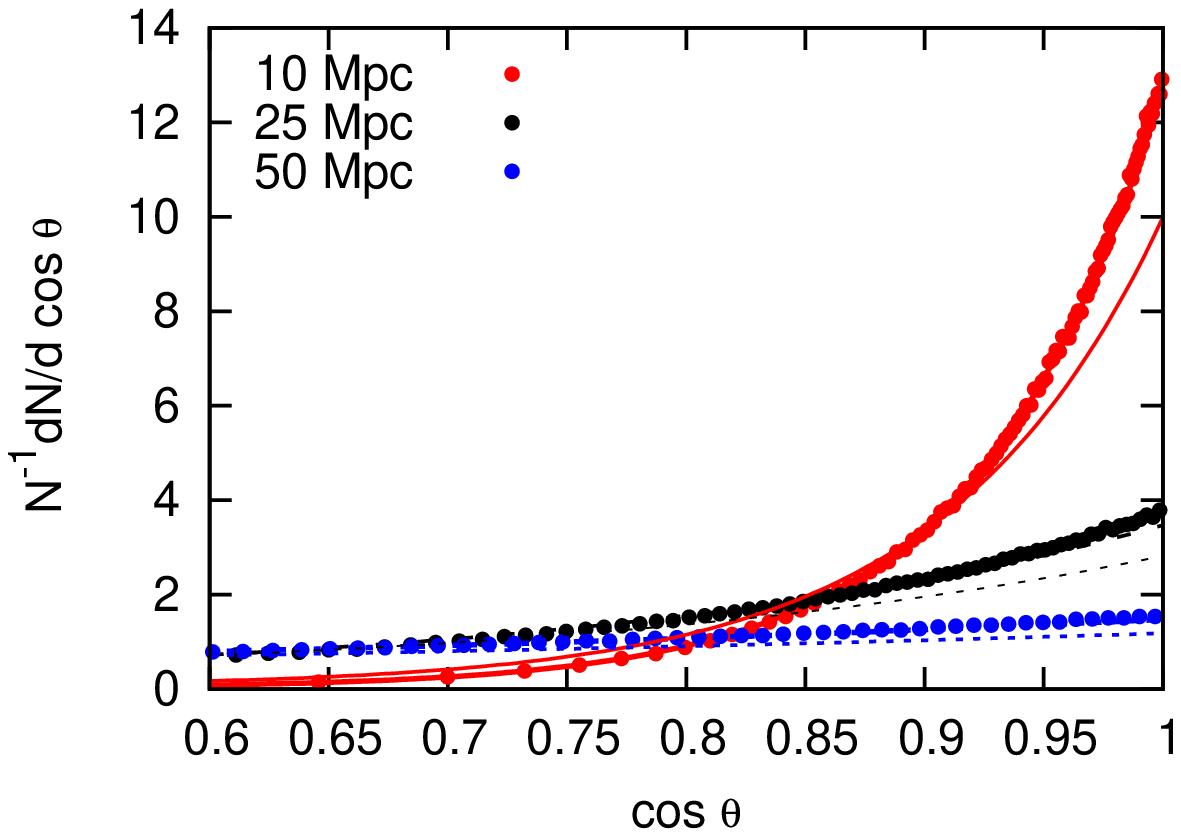}\epsfig{width=2.in,angle=-90,file=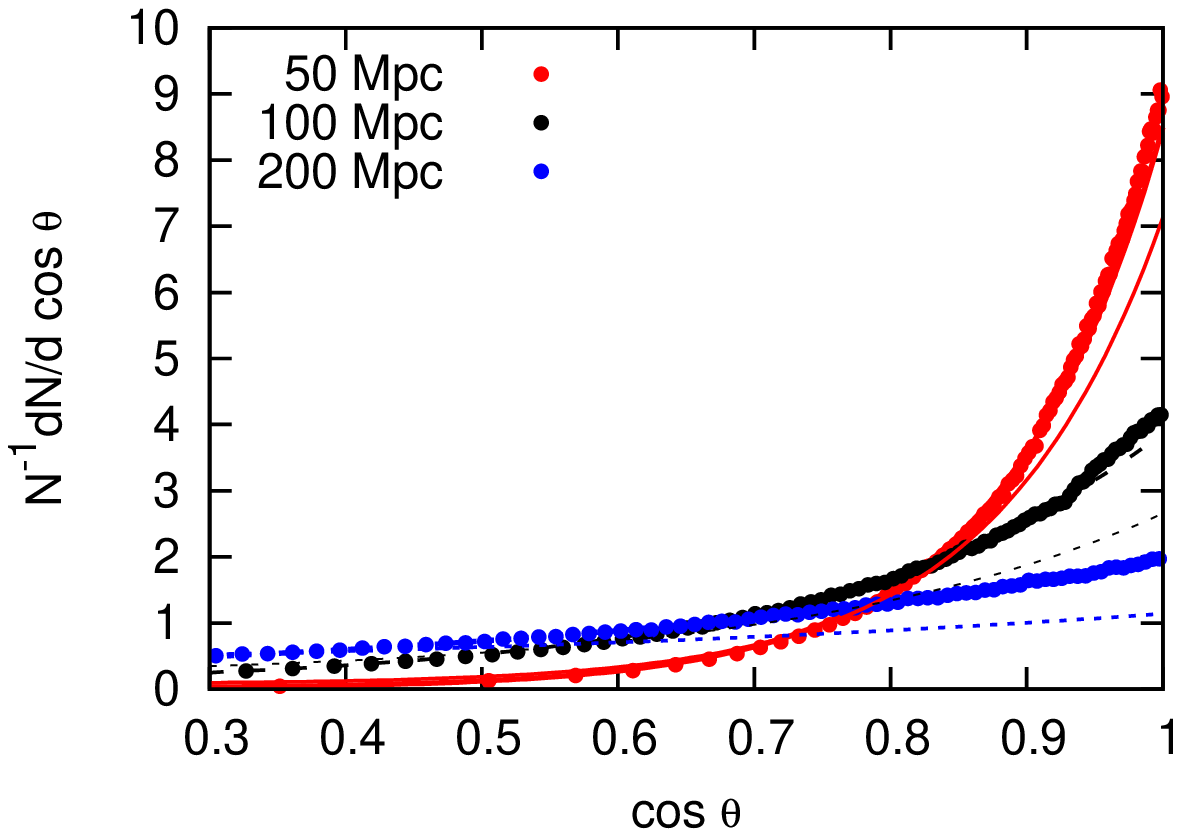}}
\vskip 1.0 truecm
\caption{Distribution of arrival directions for three values of the source distance but including energy losses. The left panel is for $E=3E_c(0)=2.7$~EeV and the right one for $E=6E_c(0)=5.4$~EeV (using $B_{rms}(0)=1$~nG and $l_c(0)=1$~Mpc). Note that $l_D(0)=39$~Mpc for $E=3E_c(0)$, while $l_D(0)=150$~Mpc for $E=6E_c(0)$. The thin lines correspond to the analytic  expressions that would be obtained without energy losses while the fits including the energy losses overlap with the simulated points.}
\label{fig:diste3}
\end{figure}

\begin{figure}[ht]
\centerline{\epsfig{width=2.in,angle=-90,file=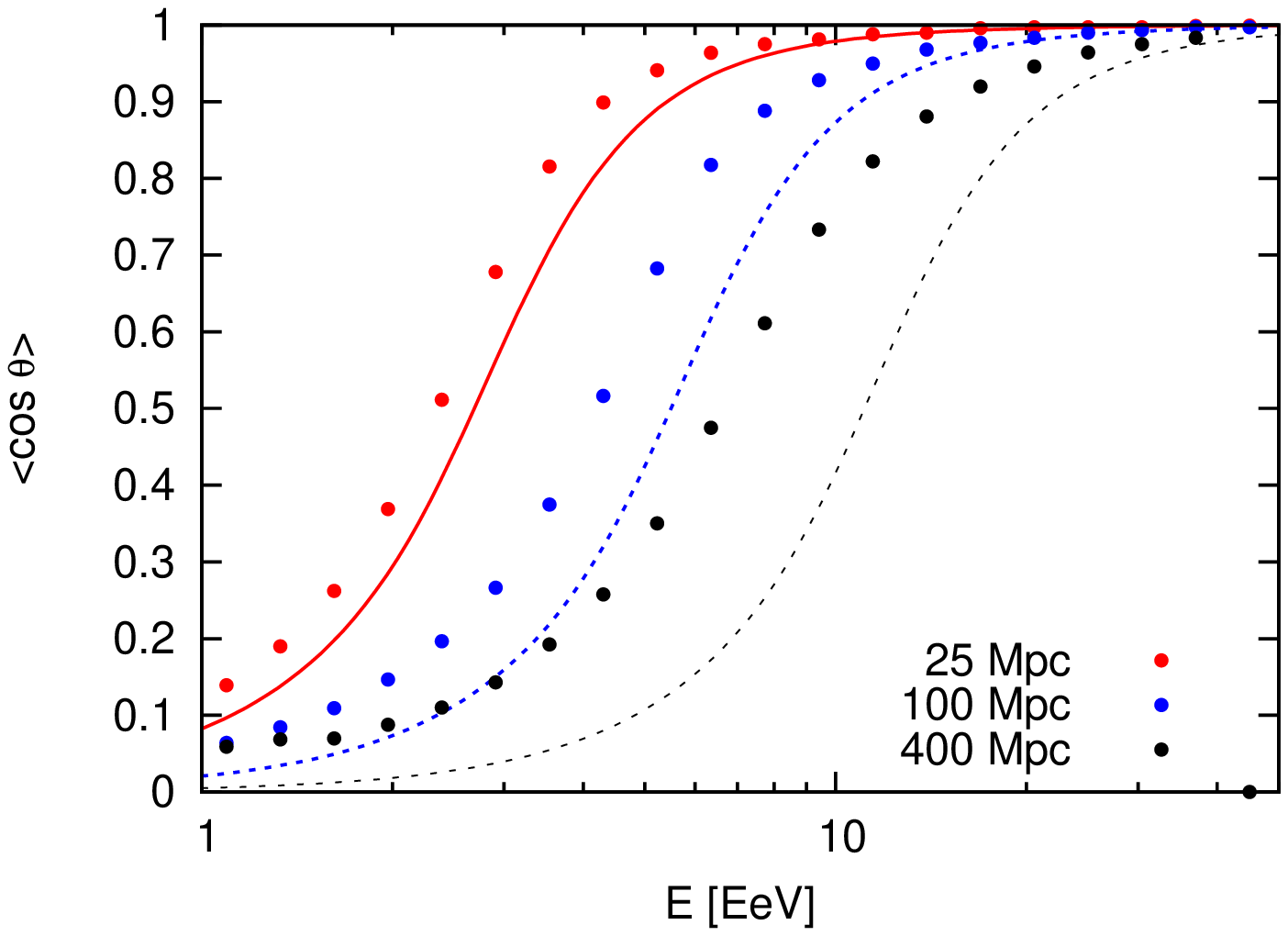}\epsfig{width=2.in,angle=-90,file=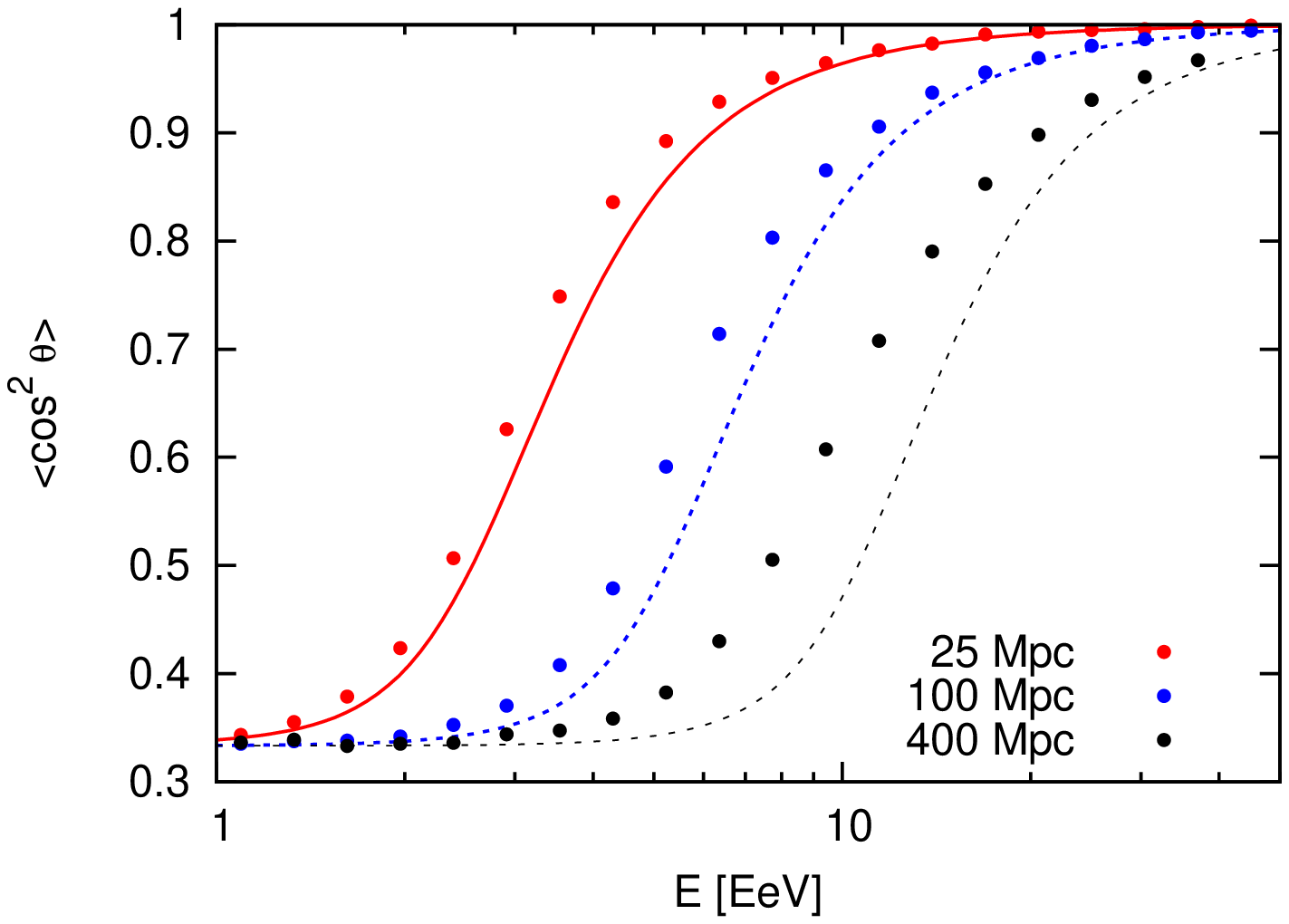}}
\vskip 1.0 truecm
\caption{Values of $\langle \cos\theta\rangle$ and  $\langle\cos^2\theta\rangle$ as a function of the arrival energy $E$ for three values of the source distance. Points include energy losses while the lines are the corresponding results in the absence of losses.} 
\label{fig:dist4}
\end{figure}

\begin{figure}[ht]
\centerline{\epsfig{width=2.in,angle=-90,file=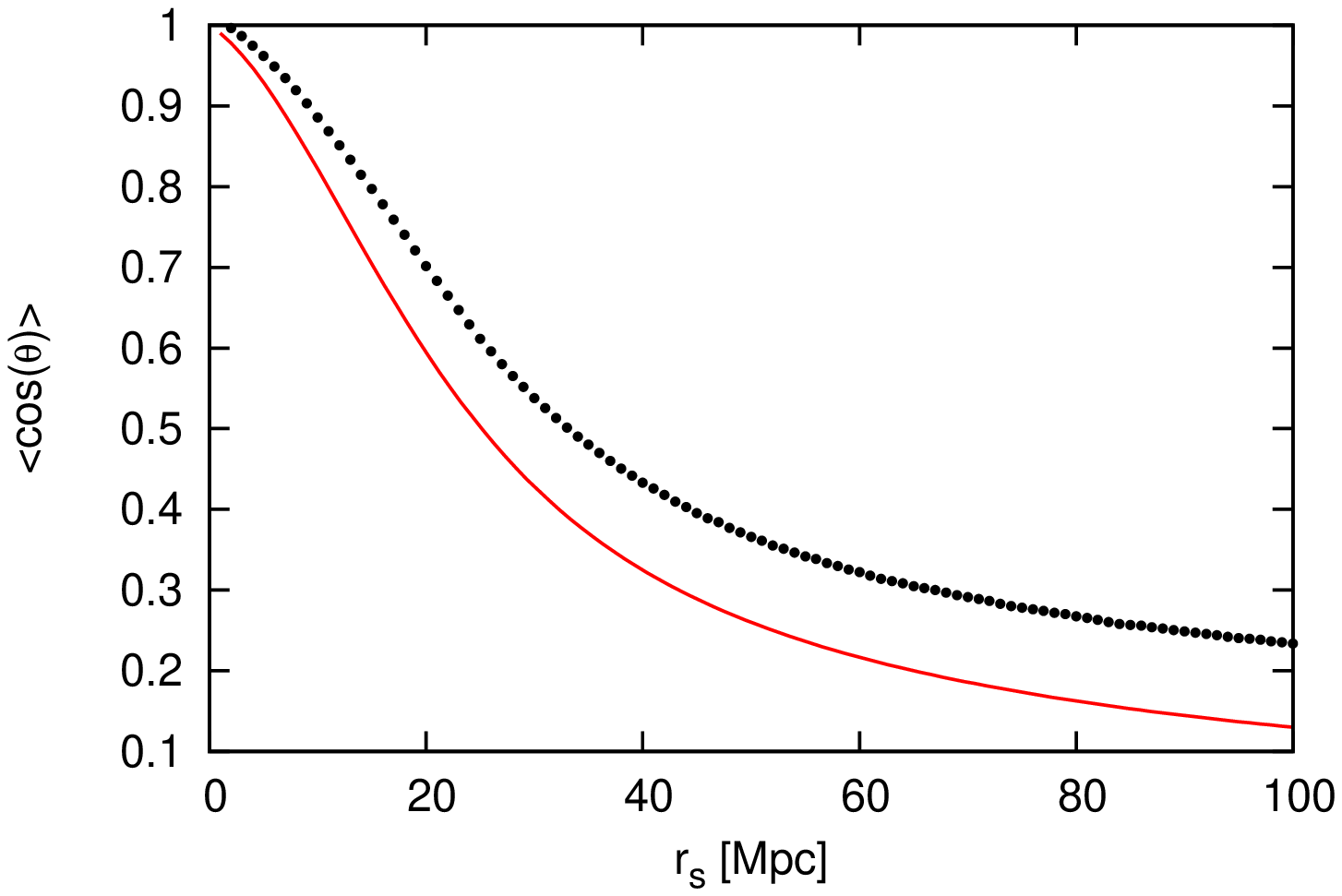}\epsfig{width=2.in,angle=-90,file=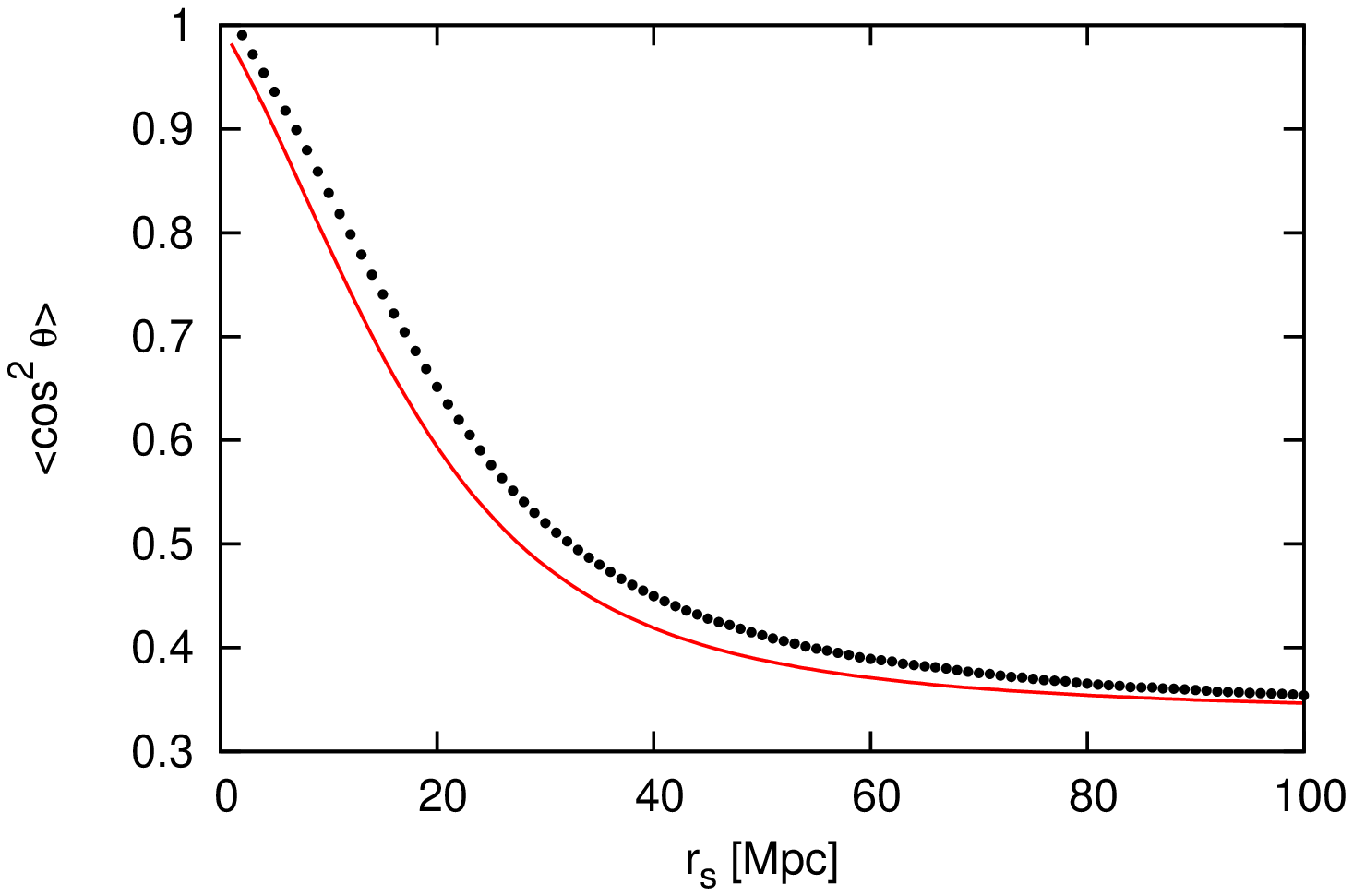}}
\centerline{\epsfig{width=2.in,angle=-90,file=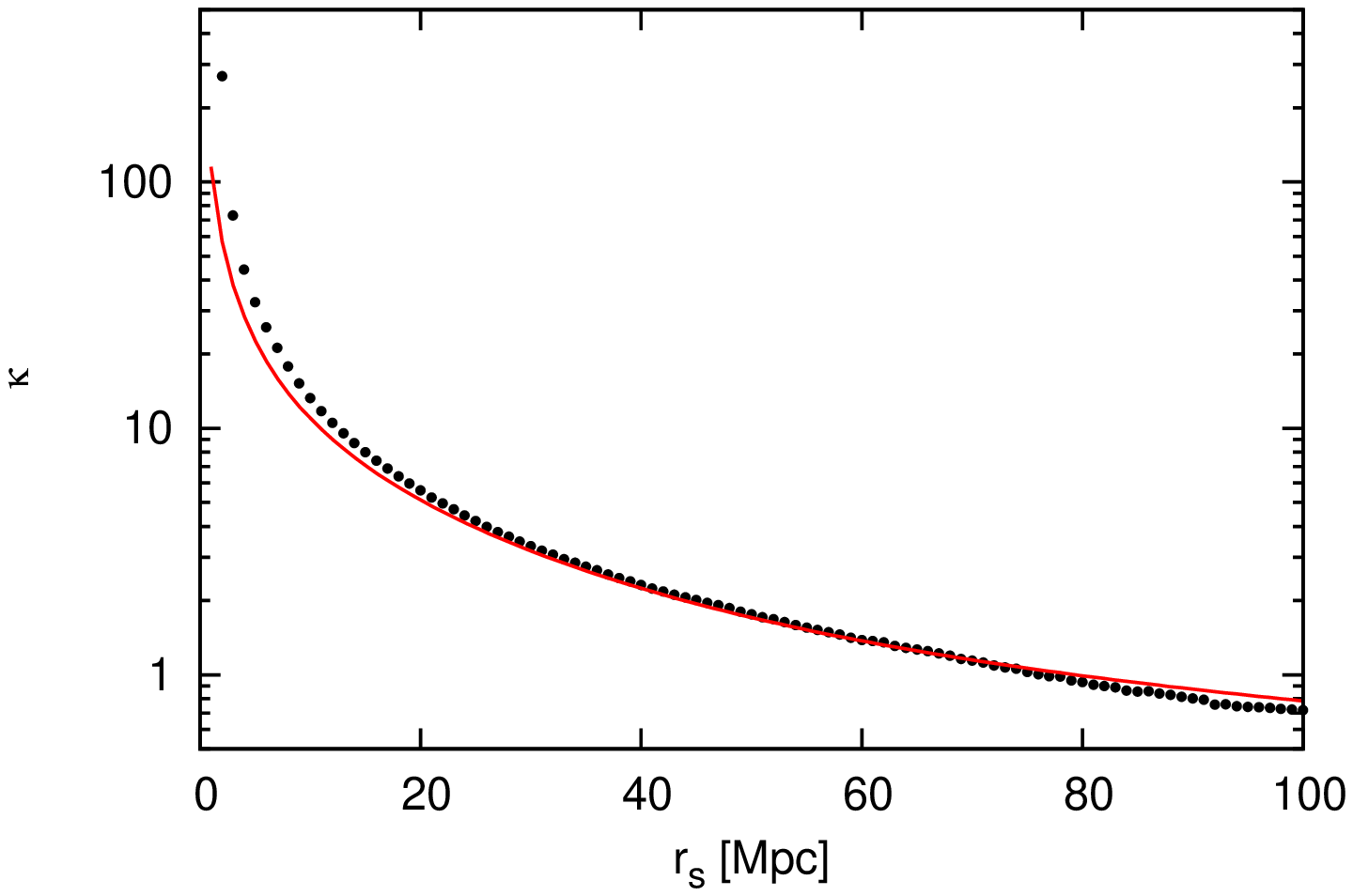}\epsfig{width=2.in,angle=-90,file=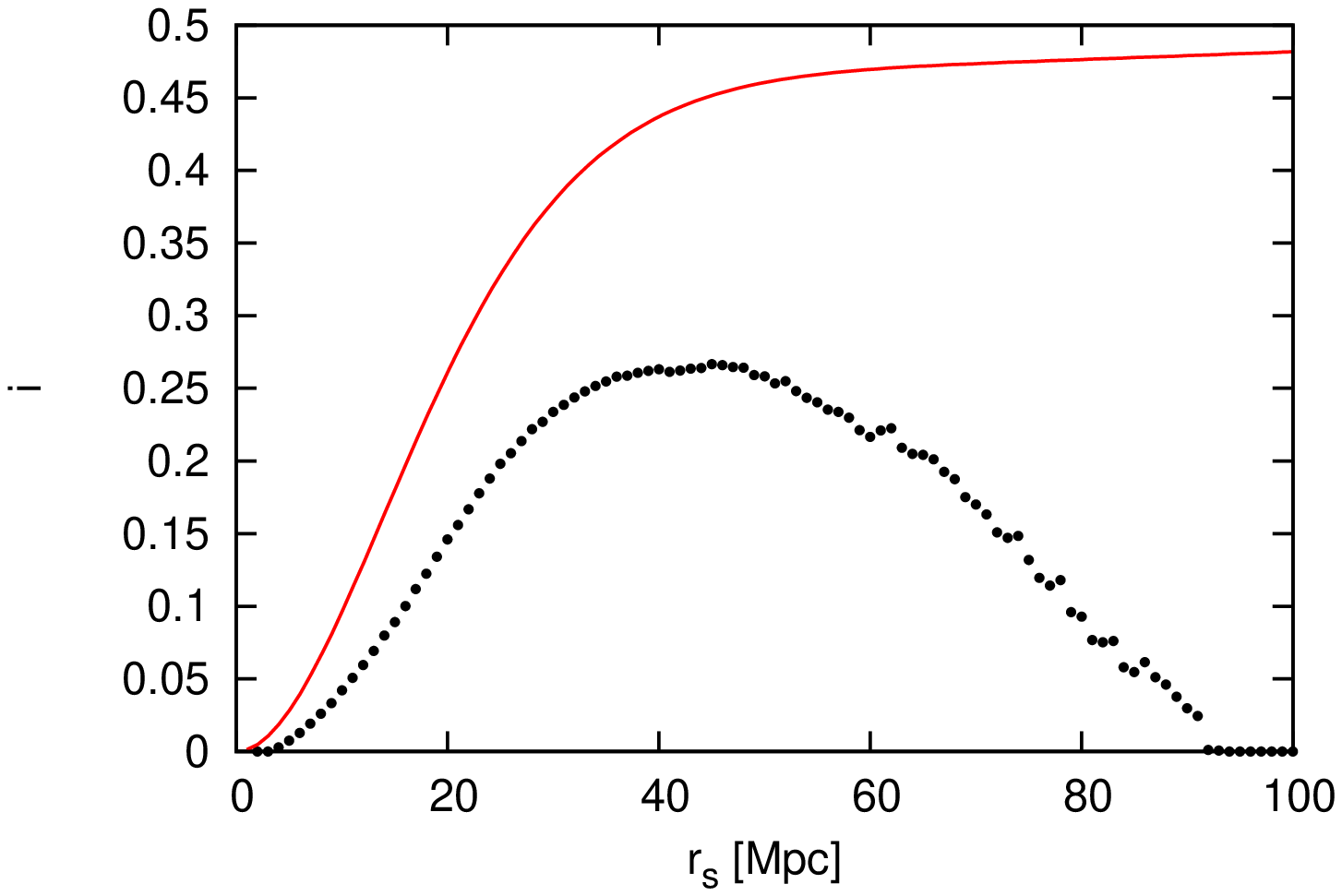}}
\vskip 1.0 truecm
\caption{Values of $\langle \cos\theta\rangle$, $\langle \cos^2\theta\rangle$, $\kappa$ and $i$ as a function of $r_s$ for a proton source and for final energies of 2.7~EeV. Dots include the effects of energy losses while  the lines neglect them.  } 
\label{paravsrs}
\end{figure}

The resulting arrival directions  of the particles can be well described with a distribution like that in Eq.~(\ref{fisher}) but with parameters somewhat different than those obtained  for a source at the same distance but in the absence of energy losses. As an example, we show in Figure \ref{fig:diste3} the distribution of the arrival directions for an observed energy $E(0)=3E_c=2.7$~EeV (left panel) and $6E_c=5.4$~EeV (right panel) and for three different values of the distance to the source.  The particles are emitted from the source with a spectrum $E^{-2}$ up to a maximum energy of $10^{21}$~eV. For each source distance we also computed the (weighted) averages $\langle \cos\theta \rangle$ and $\langle \cos^2\theta \rangle$ for all the trajectories passing through that distance. A good fit to the angular distribution is given by Eq.~(\ref{fisher}) with the parameters $\kappa$ and  $i$ obtained from Eqs.~(\ref{rk}) and (\ref{i}) with the input of $\langle \cos\theta \rangle$ and $\langle \cos^2\theta \rangle$ from the simulations. 
The resulting distributions are shown as the continuous lines in the plot and they actually overlap with the simulated points. We also show for comparison, with thin lines, the analytic result that would be obtained in the absence of energy losses that were obtained in the previous section. To understand the main differences induced by the energy losses we show in Figure~\ref{fig:dist4} the values of  $\langle\cos\theta\rangle$ and  $\langle\cos^2\theta\rangle$ as a function of the energy for different source distances and in Figure~\ref{paravsrs} we show, adopting final energies of 2.7~EeV, the values of $\langle\cos\theta\rangle$, $\langle\cos^2\theta\rangle$, $\kappa$ and $i$ as a function of $r_s$, compared to the values that would be obtained in the absence of energy losses.  We see that as a result of the losses both the values of $\langle\cos\theta\rangle$ and  $\langle\cos^2\theta\rangle$ tend to increase for a given source distance and arrival energy. It is clear that the main effect is due to the reduction of the parameter $i$, since the losses affect mostly the component resulting from long diffusion times, as is the case for the CRs contributing to the isotropic component which typically make several turns before reaching the observer. The impact of the losses in the concentration parameter $\kappa$ is moderate, and related to the fact that the particles reaching the observer with a given energy had somewhat higher energies during their propagation and this slightly enhances the values of $\kappa$ for nearby sources.  Although we focused on the case of cosmic ray protons, qualitatively similar results are obtained also for the case of cosmic ray nuclei, with the detailed behavior depending on the actual composition mixture adopted (some results are presented in the next section).

\section{Superposition of many sources}

We have considered in the previous sections the  distribution of arrival directions of UHECR protons from individual nearby sources after their propagation through the extragalactic turbulent fields. In a realistic situation, several sources at  different distances  could contribute to the observed fluxes and this would make the picture more complex.

\begin{figure}[ht]
\centerline{\epsfig{width=2.in,angle=-90,file=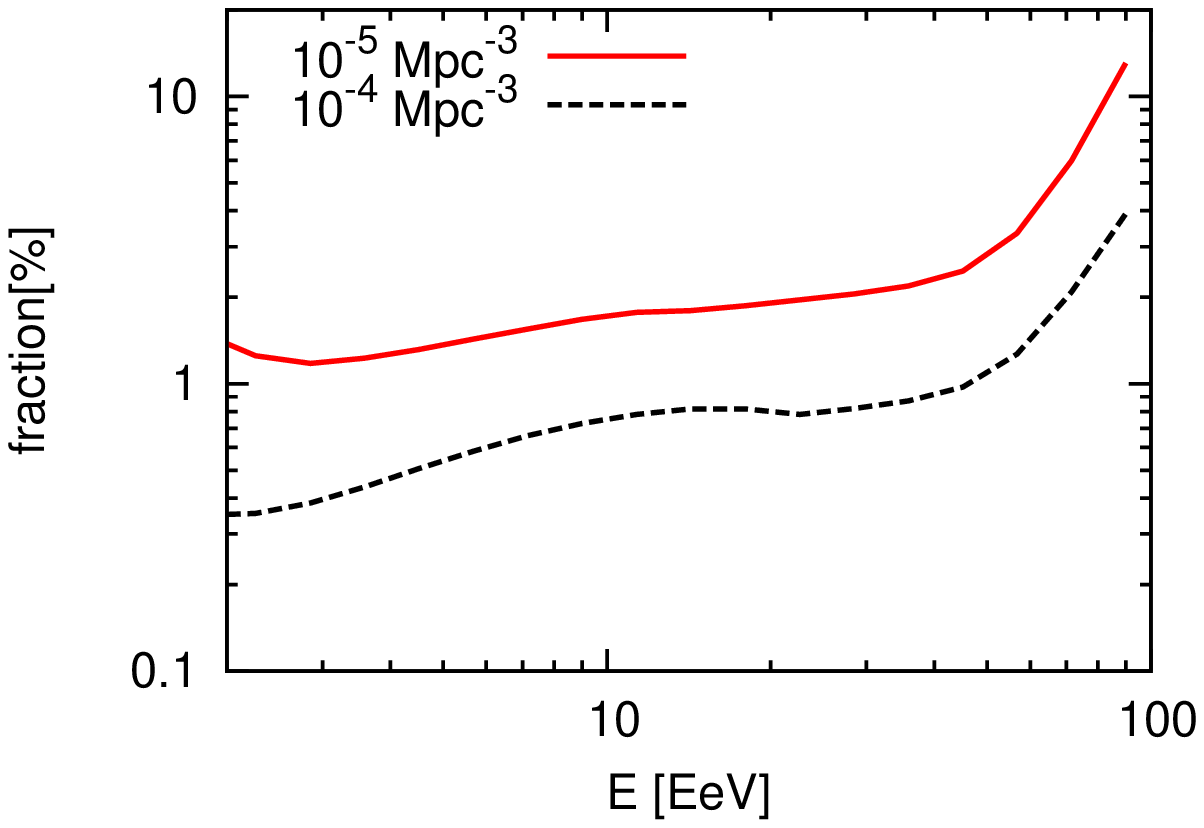}\epsfig{width=2.in,angle=-90,file=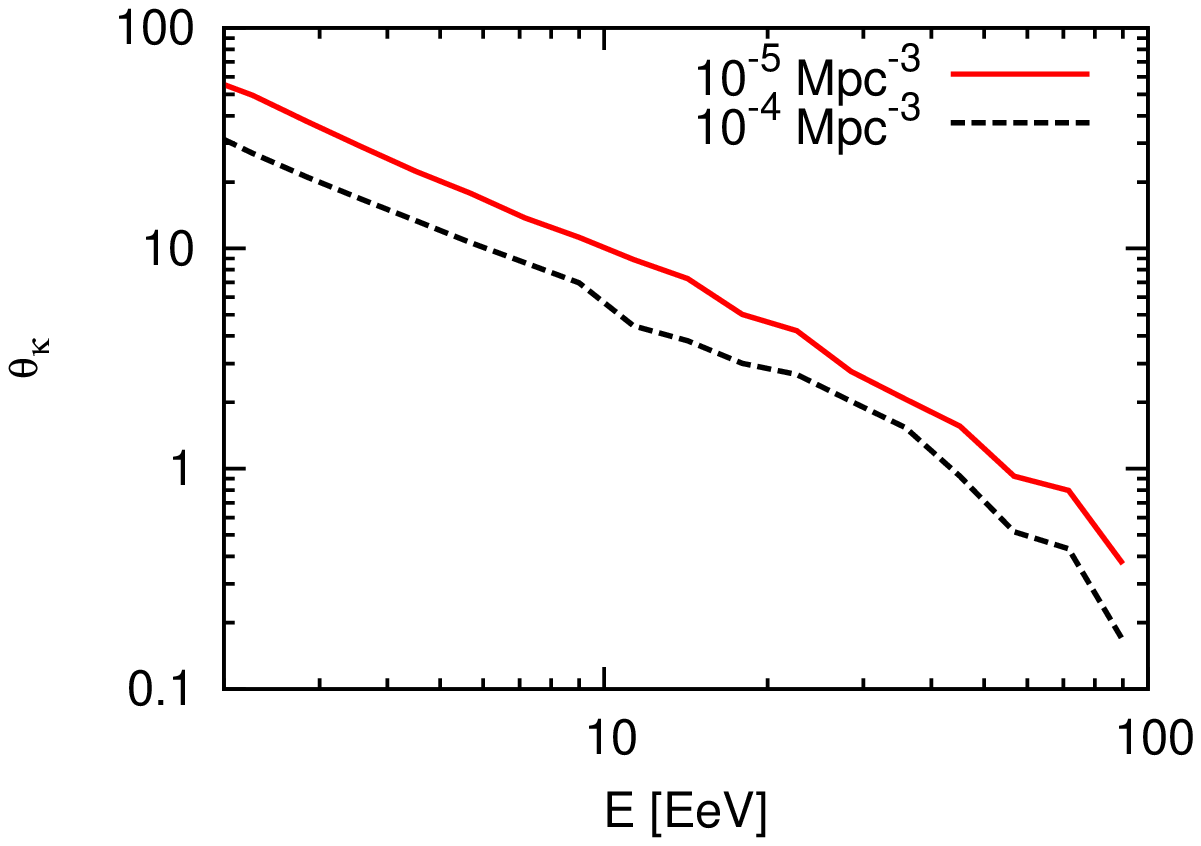}}
\centerline{\epsfig{width=2.in,angle=-90,file=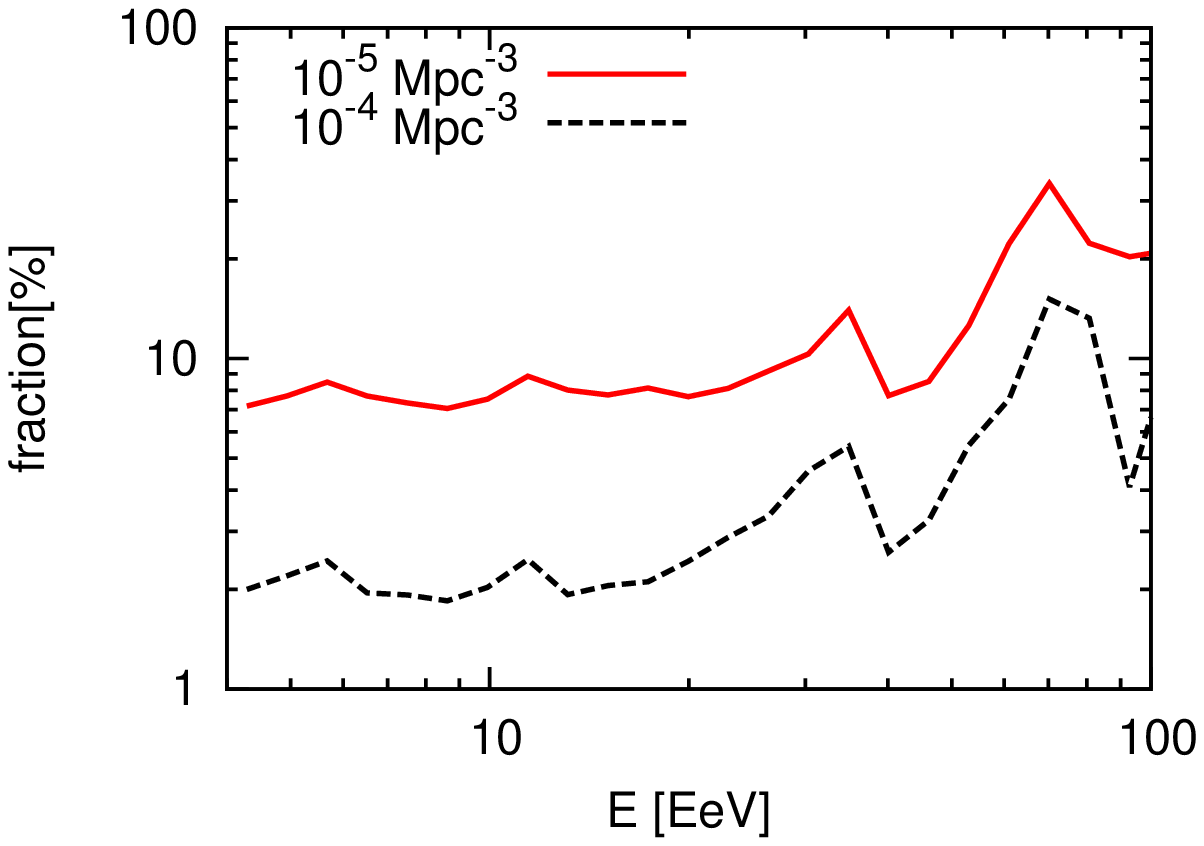}\epsfig{width=2.in,angle=-90,file=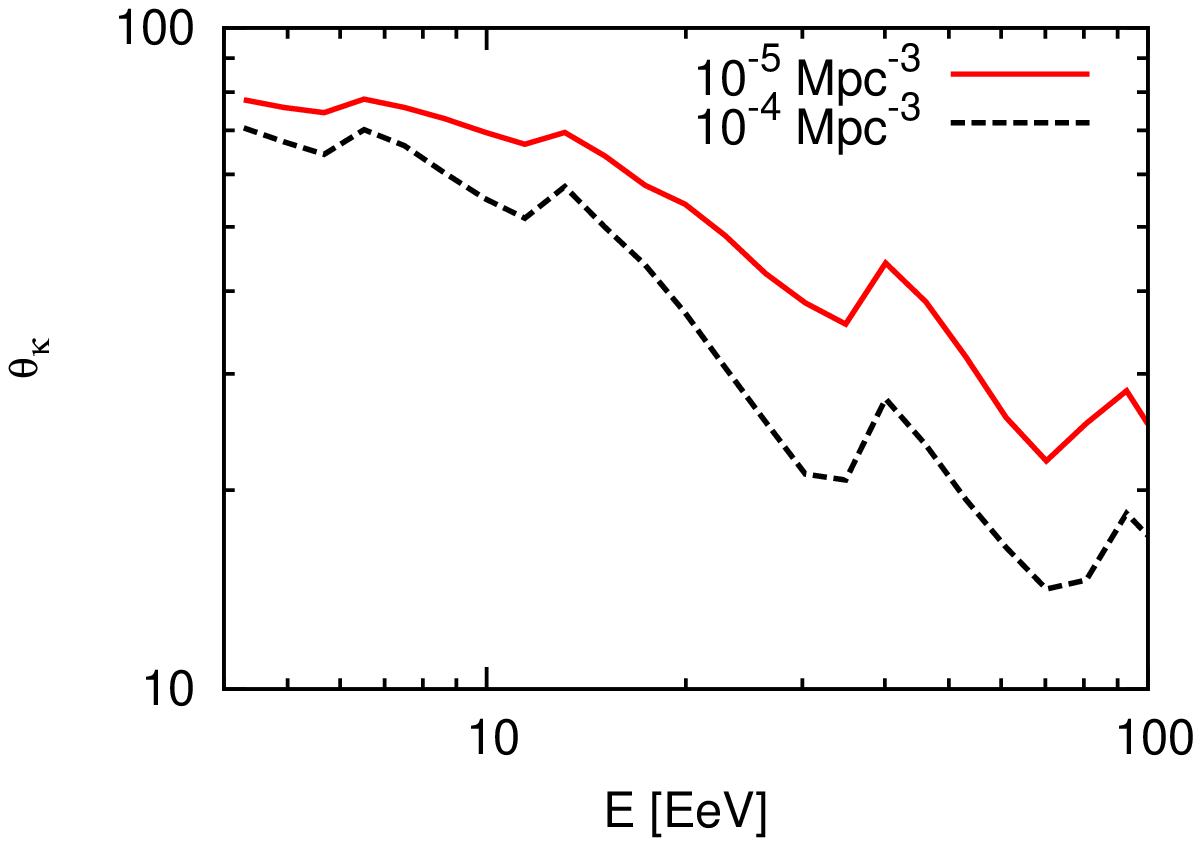}}
\caption{Left: fraction of the total flux that is expected to arrive from the closest source for two values of the source density. Right: angular spread (in degrees) of the distribution of the cosmic rays coming from the closest source. We assumed equal luminosity sources at distances equal to the corresponding expected mean values. The top panels are for proton sources while the lower ones are for nuclei adopting  fractions of p, He, N, Si and Fe nuclei as described in the text, with a maximum energy of $6Z$~EeV.} 
\label{firsts}
\end{figure}

The possibility of observing a localized excess flux of UHECRs from an individual source depends on the fraction of the total flux coming from that source and on the angular spread of the CR arrival directions. In order to estimate these quantities we will make the simplifying assumption of equal intrinsic intensity of the sources, so that for each energy the relative contribution to the flux from different sources  will only depend on the distance to the source $r_j$. For a given density of sources $\rho$, we take the radial distances from the observer to the sources as the mean expected values in an homogeneous distribution. This corresponds to adopting  for the $j$-th closest source a distance equal to the average value that is given by $\langle r_j \rangle = (3/4\pi\rho)^{1/3} \Gamma(j+1/3)/(j-1)!$. The fraction of the flux coming from the closest source can then be obtained as $f=n_1/\sum_j n_j$, with $n_j = n(\langle r_j \rangle)$. The dependence of the flux with the distance to the source can be estimated from the simulations performed, just by counting the number of particles as a function of the distance to the source  weighted by the factor $[E_g(z)/E(0)]^{-\gamma} {\rm d}E_g/{\rm d}E$. The result can be seen in Figure 6 of Paper I. The fraction of the flux from the closest source is plotted as a function of the arrival energy in the top-left panel of Figure \ref{firsts} for two values of the source density, $\rho=10^{-4}\, {\rm Mpc}^{-3}$ and $\rho=10^{-5}\, {\rm Mpc}^{-3}$. Lack of significant clustering of CRs at the highest energies constrains the source density to be larger than (0.06--5$) \times 10^{-4}\, {\rm Mpc}^{-3}$ at 95\% confidence level, depending on the magnitude of the magnetic deflections \cite{augerdensitybounds}.

 A measure of the angular spread of the CR arrival direction distribution from a given source is given by 
 $\theta_\kappa\equiv\arccos(\langle\cos\theta\rangle )$. For a source which is significantly localized, i.e. for $\kappa\gg 1$ and setting $i\simeq 0$, one finds from  Eq.~(\ref{cosq}) that 
 $\theta_\kappa\simeq\arccos(1/\tanh \kappa-1/\kappa)$, and within this angle approximately 63\% of the non-isotropic part of the flux from the source is contained. This angle is plotted in the top-right panel of Figure~\ref{firsts} as a function of the energy.

Note that in an actual observation the significance of a result will depend essentially on the ratio between the number of events from the source arriving within an angle $\theta_\kappa$ and the corresponding fluctuations expected within the same window due to the approximately isotropic background  from the remaining sources.  Ignoring for simplicity the declination dependence of the exposure of the observatory and assuming full-sky coverage, the signal to noise ratio for such observation would then be approximately given by
\begin{equation}
\frac{S}{N}\simeq \frac{0.63 f\,N_{tot}}{\sqrt{N_{tot}\Omega_\kappa/(4\pi)}}\simeq 5\left(\frac{f} {1\%}\right)\left(\frac{10^\circ}{\theta_\kappa}\right)\sqrt{\frac{N_{tot}}{5000}},
\end{equation}
where $f$ is the fraction of the flux coming from the closest source, $\Omega_\kappa$ is the solid angle subtended by the window with angular radius $\theta_\kappa$, and $N_{tot}$ is the total number of events observed above a given energy threshold. Since $\theta_\kappa\propto E^{-1}$ while below the threshold of the Greisen, Zatsepin and Kuzmin (GZK) suppression, i.e. for $E<40$~EeV, one has that $N(>E)\propto E^{-2}$ and that $f$ is approximately constant, we see that the value of $S/N$ should not depend significantly on the threshold adopted. The ratio $S/N$  should become enhanced when considering energies above the GZK threshold due to the strong suppression of the fluxes contributed by faraway sources that would lead to a steep increase in the value of $f$.
The results for $f$ in Figure~\ref{firsts} imply that for the reference parameters adopted  a significant anisotropy would be expected in scenarios of pure protons for accumulated statistics of $N_{tot} \geq 5000$ at $E>10$~EeV, which is comparable to that achieved by the Auger Observatory at present, with the significance becoming further enhanced when considering threshold energies above the GZK suppression one. The excess  would be however less significant in the case in which the deflections are larger due to magnetic fields stronger  than those adopted here, or for larger source densities.

On the other hand, for heavier nuclei the expectations are different. The corresponding results are shown in the bottom panels of Figure~\ref{firsts} and were obtained following the work in reference \cite{ha15}, adopting spectra for the different components at the sources with the same spectral index (d$N_i/{\rm d}E\propto E^{-\gamma}$) and a sharp rigidity cutoff, so that the maximum energies achieved at the source scale with the corresponding CR charges $Z$. In the example displayed we adopted $E_{max}=6Z$~EeV, a spectral index $\gamma=2$ and a mixture of 5 different CR components with relative fractions at the source (for a given energy) given by $f_p=0.19$, $f_{He}=0.19$, $f_{C}=0.4$, $f_{Si}=0.19$ and $f_{Fe}=0.03$. This set of parameters leads to an approximate qualitative agreement, after accounting for propagation effects,  with the spectrum and composition measured by the Auger Observatory \cite{augerspectrum,augercomposition1,augercomposition2}. In this case, since the composition becomes increasingly heavier beyond the proton cutoff energy of $\sim 6$~EeV, the resulting angular scale $\theta_\kappa$ remains larger than $10^\circ$. This then makes the expected significance of a localized overdensity to be smaller than in the proton case. The $S/N$  should become enhanced above the energies $\sim 40$~EeV since, due to attenuation effects during propagation the fraction of the flux contributed by the closest source becomes larger, and in this energy regime angular scales  $\theta_\kappa\sim 10^\circ$--$30^\circ$ are naturally expected.  Note that the detailed angular distribution of CRs from a source with mixed composition will actually be a superposition of different Fisher distributions (essentially one for each nuclear component) plus an isotropic term. At any given energy, the component with highest rigidity (i.e. smaller charge) will generally be the one more concentrated around the direction towards the source.

Up to now we have discussed how an extragalactic CR source would look like after accounting for the effects of the turbulent extragalactic magnetic fields. In addition to these deflections, before reaching the Earth the CRs will be further deflected by the galactic magnetic fields, mostly by the regular component  (the random component leads to deflections which are sub-dominant with respect to the extragalactic ones already considered if the strength of the extragalactic turbulent field is indeed ${\cal O}$(nG)).
The regular galactic magnetic field $B_{reg}$ will give rise to coherent energy dependent deflections which would displace and elongate the source images in a direction orthogonal to the magnetic field direction. The typical deflections induced are
\begin{equation}
\delta_{reg}\simeq 25^\circ \frac{6\,\text{EeV}}{E/Z}\left|\int\frac{\text{d}\vec{s}}{\text{kpc}}\times\frac{\vec{B}_{reg}}{3\,\mu\text{G}}\right|,
\end{equation}
where d$\vec{s}$ is the displacement along the CR trajectory and the typical strength of the galactic fields is a few $\mu$G, although the actual values and detailed overall geometries are still uncertain, with some recent models being \cite{JF,PT}. The resulting deflections depend on the direction considered and can be relatively larger in directions close to the galactic center and generally weaker in directions orthogonal to the galactic plane. In the case of proton scenarios these deflections would amount to only few degrees at the highest energies while in scenarios having mixed composition and with maximum rigidities $<10$~EV they could be of few tens of degrees even for rigidities close to the maximum value allowed. 
These deflections can affect the ability to reconstruct the actual source directions, especially if a heavy CR component is dominant, but on the other hand they also provide a handle to learn about the galactic magnetic field and the CR charges involved.

\section{Lensing effects on diffuse sources}\label{lensing}
Besides the coherent deflections and the diffusive behaviors previously discussed,  there are different situations in which magnetic fields can also produce lensing effects upon the UHECR fluxes (see e.g. \cite{ha99}). For turbulent magnetic fields  these effects can appear when the typical transverse displacement of the CR trajectories with respect to the straight line propagation reaches values comparable to the coherence length of the magnetic field \cite{wa96,ha02}. In this case CRs can travel through different paths  from the source to the observer and hence multiple images of the source appear. For a given CR energy this condition is met  when the typical deflection satisfies $\theta\simeq \sqrt{r_s/l_D}\simeq l_c/r_s$, i.e. for $r_s/l_D\simeq (l_c/l_D)^{2/3}$. This corresponds to an energy $E\simeq E_c(r_s/l_c)^{3/2}/2$, which can be quite large if $r_s\gg l_c$. Very strong magnifications of the fluxes can be produced when multiple images appear, with a strength which is energy-dependent. In this regime the results of the diffusive method obtained in previous sections describe the average over different field realizations, or over several source positions for a given field realization, but are not a realistic description for individual sources in a given  realization. At smaller energies, or for larger source distances at a given energy, the number of images increases significantly and one reaches the regime in which the diffusive treatment applied in previous sections holds also for individual sources.  

Other interesting lensing effects are those due to the regular component of the galactic magnetic field. Since this field is coherent over kpc scales and is present quite close to us, multiple imaging requires relatively large deflections and can hence only take place at lower energies (typically for $E/Z<20$~EeV, see e.g. \cite{ha00}). To describe these phenomena it is useful to consider the mapping between the sky coordinates of the CR arrival directions at the observer location (image plane) and the CR directions before entering the galactic magnetic field (source plane), taken for instance at a sphere of 30~kpc radius around the observer. We refer to this sphere as the source plane even if the CRs from the actual sources have already diffused through the extragalactic field before reaching it, and hence the sources in this plane will already look as extended objects. When multiple images are present this mapping is multiple valued and the singularities that appear can be described in terms of cusps and folds in the mapping, generically called caustics, while the corresponding directions in the image plane are called critical lines. The most ubiquitous feature is that of folds, which are lines across the source plane indicating the directions at which new images appear. These images appear in pairs and they are strongly magnified near the fold, leading actually to a divergent magnification in the limit of point-like sources when the fold overlaps with the source direction. 

Adopting angular coordinates parallel, $\theta_{\parallel}$, and perpendicular, $\theta_\perp$, to the fold, the magnification, defined as the ratio between the flux resulting in the presence of the magnetic field to the one that would be expected when no magnetic fields are present, takes the following approximate form near a fold caustic \cite{ha00} 
\begin{equation}
A({\bf\theta})\simeq A_c+A_{div}({\bf\theta}),
\end{equation}
with $A_c$ constant and 
\begin{equation}
A_{div}({\bf\theta})=\frac{A_f}{\sqrt{\theta_\perp-\theta_f}}H(\theta_\perp-\theta_f),
\label{adiv}
\end{equation}
with $\theta_f$ indicating the location of the fold in the  $\theta_\perp$ direction and $H$ being the Heaviside function. It is also convenient to express the constant $A_f$ as $A_f\equiv\sqrt\Delta$, where now $\Delta$ is the angular distance from the fold for which $A_{div}=1$. The actual values of $\Delta$ depend on the energy considered, and simulations with realistic galactic magnetic field models show that values from few to few tens of degrees are typical once folds form and before entering the low-energy diffusive regime of the propagation through the  galactic magnetic field.

We want to discuss now what is the effect of the presence of a fold, produced for instance by the regular magnetic field, in the magnification of a point source that has been spread by the action of a turbulent extragalactic magnetic field, assuming that this spreading is moderate (say less than 30$^\circ$) so that the previous expression for the magnification is reasonable.
The effect of the constant term $A_c$ is the same for a point source or a spread one, and is just to trivially amplify the flux from the source by that factor. We hence focus on the effects of the term $A_{div}$ alone, and evaluate how the divergence in the magnification is smoothed out when averaged over a moderately spread source.  We will take the origin of the angular coordinates at the fold location, so that the fold is at $\theta_f=0$, extending towards $\theta_f>0$, and such that the center of the source image is at $\theta_\parallel=0$ and $\theta_\perp=\theta_s$. 

As shown in the previous sections the image of the source in this plane will be spread (assuming for simplicity just the expression for a single mass component) according to
\begin{equation}
\frac{1}{N}\frac{{\rm d}N}{{\rm d}\,\cos\theta}=\frac{i}{2}+(1-i)\frac{\kappa\exp(\kappa\cos\theta)}{2\,{\rm sinh}\,\kappa}\equiv F(\cos\theta,i,\kappa),
\end{equation}
where  $\kappa$ determines the spread of the source image due to the turbulent field.

We will  introduce dimensionless variables $x\equiv \theta_\perp/\theta_\kappa$ and $y\equiv \theta_\parallel/\theta_\kappa$, where the angle  $\theta_\kappa=\text{arccos}(\langle \cos\theta\rangle )\simeq {\rm arccos}(1/\text{tanh}\kappa-1/\kappa)$ characterizes the angular width of the distribution of the CRs after having traversed the  turbulent extragalactic magnetic field.
We then get for the magnification $A_{div}$ averaged over the source image 
\begin{equation}
\langle A_{div}\rangle(x_s)\simeq \frac{\theta_\kappa^2}{2\pi}\sqrt{\frac{\Delta}{\theta_\kappa}}\int{\rm d}x\,{\rm d}y\,F(\cos(\theta_\kappa\sqrt{(x-x_s)^2+y^2)}),i,\kappa)\frac{H(x)}{\sqrt{x}},
\label{adivav}
\end{equation}
where $x_s\equiv \theta_s/\theta_\kappa$. The integration here should extend so as to cover angles of a few times $\theta_\kappa$ around the source direction (assuming the expression of the fold magnification to be valid in that range). 

\begin{figure}[t]
\centerline{\epsfig{width=3.in,angle=0,file=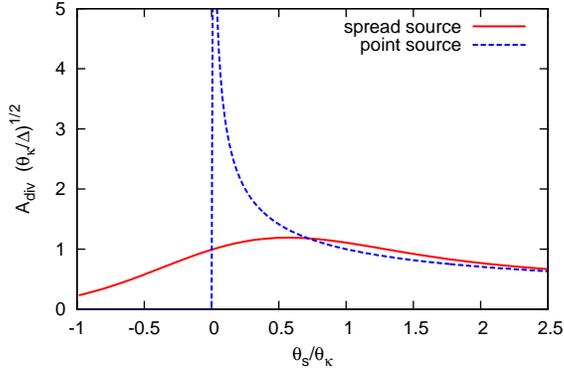}}
\vskip 1.0 truecm
\caption{Average magnification, multiplied by $\sqrt{\theta_\kappa/\Delta}$, as a function of the normalized angular distance from the source to the fold.} 
\label{fmag}
\end{figure}

 Note that in order to have relatively small spreads, e.g. $\theta_\kappa<30^\circ$,  the source distance should be much smaller than the diffusion length, i.e.  $R\leq 0.2$ (see Figure~\ref{fig:cos}), and this implies that $\kappa$ has to be large, $\kappa\simeq 3/R\geq 15$ (see Figure~\ref{fig:ki}). In this limit one has that $i\simeq 2R^2$ so that to a good approximation one may set $i=0$ (moreover, energy loss effects will further suppress the value of $i$). In the Figure~\ref{fmag} we show the quantity $\langle A_{div}\rangle\times \sqrt{\theta_\kappa/\Delta}$ as a function of $x_s$, as well as the result for a point source resulting from Eq.~(\ref{adiv}).
For the case of the smeared source  we adopted $\kappa=20$, but actually the result is essentially independent from $\kappa$ as long as $\kappa\gg 1$.  Indeed, for large $\kappa$ one has ${\rm coth}\,\kappa\simeq 1-2\exp(-2\kappa)\simeq 1$, and hence one gets $\cos\theta_\kappa\simeq 1-1/\kappa$, so that $\theta_\kappa^2\simeq 2/\kappa$. In this case one obtains that
\begin{equation}
\langle A_{div}\rangle \sqrt{\frac{\theta_\kappa}{\Delta}}\simeq \frac{1}{\sqrt{\pi}}\int_{-x_s}^\infty {\rm d}x\,\frac{\exp(-x^2)}{\sqrt{x+x_s}},
\end{equation}
where we extended the integral to infinity since the integrand is exponentially suppressed. We see then that the result is clearly independent of $\kappa$.
As can be observed in Figure~\ref{fmag}, once the spread of the source is taken into account the divergent behavior of the magnification near the fold caustic is smoothed out, and typical values $\langle A_{div}\rangle\simeq \sqrt{\Delta/\theta_\kappa}$ are obtained for $|\theta_s-\theta_\kappa|\leq\theta_\kappa$ \footnote{We have not accounted in this calculation for the angular resolution of cosmic ray experiments, typically of the order of $1^\circ$, that induce an additional smearing, since its effect is expected to be small compared to that of the spread of the source.}. These magnifications are still large if the spread $\theta_\kappa$ due to the turbulent field  is much smaller than the fold width $\Delta$, but otherwise the magnification near the fold can be significantly reduced\footnote{In ref.~\cite{ha00} it was shown that the divergent magnification produced by a fold at the energy at which it crosses the location of a point source is also smoothed out once the flux in a finite energy bin is considered.}. A similar behavior is known to happen in the analogous case of gravitational lensing once the finite size of the sources is taken into account. 

\section{Discussion}
We have considered in this paper the arrival direction distribution of UHECRs from individual nearby sources after their propagation through an extragalactic turbulent magnetic field. We have found that the CR angular distribution can be accurately described by the superposition of a Fisher distribution, which generalizes the Gaussian one to the sphere, plus an isotropic component which is required to account for part of the CRs that diffused for very long times. When the source distance is much larger than the diffusion length, the distribution of arrival directions is characterized by a dipolar term, as was obtained in refs.~\cite{I,ha15}, and we further obtained here the quadrupolar contribution to the distribution. For source distances not much larger than the diffusion length, a localized excess around the source direction results, having an approximately Gaussian distribution with an energy dependent angular extent.

 There are actually some hints at  present suggesting possible overdense localized regions at the highest cosmic ray energies, i.e. a hot spot on angular scales of $20^\circ$ reported by the Telescope Array Collaboration for $E>57$~EeV \cite{tahs} and an excess on a $12^\circ$ radius region not far from the direction towards Centaurus~A reported by the Auger Collaboration for $E>58$~EeV \cite{auhs}. If this angular spread were due to the effects of the turbulent extragalactic magnetic fields one would infer that the CRs produced by the individual sources should lead to  $\theta_\kappa\simeq 10^\circ$--$20^\circ$ for $E\sim E_{th}\simeq 60$~EeV. The size of these deflections is consistent with the results obtained in this work for the mixed composition scenario (see figure~\ref{firsts}). Since for large $\kappa$ one has $\theta_\kappa\simeq \sqrt{2/\kappa}\simeq \sqrt{2R/3}$, this would lead to $R\simeq 1.5\,\theta_\kappa^2$
and hence the source distance should be a small fraction of the diffusion length. Using that for $E\gg E_c$ one has that $l_D\simeq 4l_c(E/E_c)^2$, this would suggest that the source distance should be 
$r_s\simeq 20\,\text{Mpc}(10/Z)^2(\text{nG}/B)^2(\text{Mpc}/l_c)(\theta_\kappa/15^\circ)^2$, with $Z$ being the dominant CR charge at these energies.

If future CR data were to provide more significant evidence of individual UHECR sources as well as information on the CR composition, detailed analyses along the lines introduced in this work would enhance our understanding of the intervening magnetic fields and of the source properties.

\section*{Acknowledgments}
 Work supported by CONICET and ANPCyT, Argentina.


\begin{thebibliography}{99}

\bibitem{auhs} A. Aab et al. (Pierre Auger Collaboration), Astrophys. J. {\bf 804} (2015) 15.

\bibitem{audip}  A. Aab et al. (Pierre Auger Collaboration), Astrophys. J. {\bf 802} (2015) 111.

\bibitem{tahs} R.U. Abassi et al. (Telescope Array Collaboration), Astrophys. J. {\bf 790} (2014) L21.

\bibitem{I}D. Harari, S. Mollerach and E. Roulet, Phys. Rev. D {\bf 89} (2014) 123001.

\bibitem{ha15} D. Harari, S. Mollerach and E. Roulet, Phys. Rev. D {\bf 92} (2015) 063014.

\bibitem{ac99} A. Achterberg, Y.A. Gallant, C.A. Norman and D.B. Melrose, astro-ph/9907060.

\bibitem{fi87}N. I. Fisher, T. Lewis and B. J. J. Embleton, ``Statistical Analysis of Spherical Data", Cambridge
Univ. Press, Cambridge (1987).

\bibitem{be07} V. Berezinsky and A.~Z. Gazizov, 
Astrophys. J. {\bf  669} (2007) 684.

\bibitem{augerdensitybounds}P. Abreu et al. (Pierre Auger Collaboration), JCAP  {\bf 05} (2013) 009. 

\bibitem{augerspectrum}I. Vali\~no, for the Pierre Auger Collaboration, $34^{th}$ ICRC, The Hague, Netherlands (2015) [arXiv:1509.03732]

\bibitem{augercomposition1} A. Aab et al. (Pierre Auger Collaboration), Phys. Rev. D {\bf 90} (2014) 122005. 

\bibitem{augercomposition2} A. Aab et al. (Pierre Auger Collaboration), Phys. Rev. D {\bf 90} (2014) 122006. 

\bibitem{JF} R. Jansson and G. Farrar, Astrophys. J. {\bf 757} (2012) 14.

\bibitem{PT} M.S. Pshirkov et al., Astrophys. J. {\bf 738} (2011) 192.

\bibitem{ha99}D. Harari, S. Mollerach and E. Roulet, JHEP {\bf 08} (1999) 022.

\bibitem{wa96}  E. Waxman and J. Miralda-Escud\'e, Astrophys. J. {\bf 472} (1996) L89; J. Miralda-Escud\'e
and E. Waxman, Astrophys. J. {\bf 462} (1996) L59.

\bibitem{ha02} D. Harari, S. Mollerach, E. Roulet and F. S\'anchez, JHEP {\bf 0203} (2002) 045.

\bibitem{ha00}D.~Harari, S.~Mollerach and E.~Roulet, JHEP {\bf 02} (2000) 035.

\end{thebibliography}
\end{document}